\documentclass[sigconf]{acmart}

\renewcommand\footnotetextcopyrightpermission[1]{} % removes footnote with conference information in first column
\usepackage{booktabs}
\usepackage{multirow}
\usepackage{xspace}

\usepackage{enumitem}

\newcommand{\ie}{\emph{i.e.,}\xspace}
\newcommand{\ignore}[1]{}

\newcommand{\wrt}{w.r.t.\xspace}

\AtBeginDocument{%
  \providecommand\BibTeX{{%
    \normalfont B\kern-0.5em{\scshape i\kern-0.25em b}\kern-0.8em\TeX}}}

\acmConference[WSDM Cup '22]{Make sure to enter the correct
  conference title from your rights confirmation emai}{February 21--25,
  2022}{Arizona, USA}
\acmPrice{15.00}
\acmISBN{978-1-4503-XXXX-X/18/06}

\begin{document}

%\title{AntGraph: Leveraging Multi-hop Graph Representation for Temporal Link Prediction}
\title{An Effective Graph Learning based Approach for Temporal Link Prediction: The First Place of WSDM  Cup 2022}
% \subtitle{The 1st place solution for Temporal Link Prediction at WSDM Cup 2022}

\author{Qian Zhao}
\affiliation{%
  \institution{Ant Group}
  \country{}
}
\email{zq317110@alibaba-inc.com}
\author{Shuo Yang}
\affiliation{%
  \institution{Ant Group}
  \country{}
}
\email{kexi.ys@antfin.com}
\author{Binbin Hu$^*$}
\affiliation{%
  \institution{Ant Group}
  \country{}
} \thanks{$^*$Corresponding author}
\email{bin.hbb@antfin.com}
\author{Zhiqiang Zhang}
\affiliation{%
  \institution{Ant Group}
  \country{}
}
\email{lingyao.zzq@alipay.com}
\author{Yakun Wang}
\affiliation{%
  \institution{Ant Group}
  \country{}
}
\email{feika.wyk@antgroup.com}
\author{Yusong Chen}
\affiliation{%
  \institution{Ant Group}
  \country{}
}
\email{chenyusong.cys@alibaba-inc.com}
\author{Jun Zhou}
\affiliation{%
  \institution{Ant Group}
  \country{}
}
\email{jun.zhoujun@alibaba-inc.com}
\author{Chuan Shi}
\affiliation{%
  \institution{Beijing University of Posts and Telecommunications}
  \country{}
}
\email{shichuan@bupt.edu.cn}


\begin{abstract}

Temporal link prediction, as one of the most crucial work in temporal graphs, has attracted lots of attention from the research area. The WSDM Cup 2022 seeks for solutions that predict the existence probabilities of edges within time spans over temporal graph.
This paper introduces the solution of {\itshape AntGraph}, which wins the 1st place in the competition. 
We first analysis the theoretical upper-bound of the performance by removing temporal information, which implies that only structure and attribute information on the graph could achieve great performance. Based on this hypothesis, then we introduce several well-designed features. Finally, experiments conducted on the competition datasets show the superiority of our proposal, which achieved AUC score of 0.666 on dataset A and 0.902 on dataset B, the ablation studies also prove the efficiency of each feature.
\end{abstract}





\maketitle

\section{Introduction}
% 1. 随着 ****，什么变得越来越重要。。。
% 2. 本次比赛提出了什么任务，该任务具体做什么。
% 3. 本方案大致是怎么解决的 fig 1，最后取得了什么效果。
\begin{table}[t]
  \centering
  \caption{The statistics of two datasets. Note that ``\# Node'' of Dataset B is obtained by the maximum value of node ids, and ``Inter.'' is short for ``Intermediate''.}
  \label{tab:description_of_datasets}
  \setlength{\tabcolsep}{5.7mm}{
  \begin{tabular}{cccc}
  \toprule
  {} & {Dataset A} & {Dataset B} \\
  \midrule
  {\# Train} & {27, 045, 268} & {8, 278, 431} \\
  {\# Initial Test} & {8, 197} & {3, 863} \\
  {\# Inter. Test} & {49, 903} & {49, 940} \\
  {\# Final Test} & {200, 000} & {200, 000} \\
  \midrule
  \midrule
  {\# Nodes}  & {19, 942} & {1, 304, 045} \\
  {\# Edges}  & {27, 045, 268} & {8, 278, 431} \\
  {\# Node feat.}  & {8} & {N.A.} \\
  {\# Edge feat.}  & {N.A.} & {768} \\
  {\# Edge type } & {248} & {14} \\
  \bottomrule
  \end{tabular}}
  \end{table}
  
As graphs are ubiquitously exist in a wide range of real-world applications, many problems can be formulated as specific tasks over graphs. And \emph{link prediction}~\cite{kumar2020link}, as one of the most important task in graph-structured datasets, is widely applied in biology~\cite{turki2017link}, recommendation~\cite{ying2018graph,hu2018leveraging} and finance~\cite{yang2021inductive}.
Meanwhile, real-world data is usually evolving over time, some following literature~\cite{yang2020financial,wang2021apan} consider devising temporal graph learning models to uncover temporal information. However, predicting the links on a temporal graph is more non-trivial.
WSDM Cup 2022 calls for solutions that predicting the probability of a link within a period of time. In this paper, we will introduce the solution of \emph{AntGraph} team, which ranks the first of the competition (achieved AUC score of 0.666 on dataset A and 0.902 on dataset B).
And this technical report is organized as following:
\begin{itemize}[leftmargin=*]
    \item First, we give some statistics on the datasets, do some exploratory analyses and introduce the motivation of the method. According to the data analyses, we surprisingly find that removing the time span information in prediction could also achieve satisfactory performance.
    
    \item Subsequently, we introduce the data processing flow, enumerate several feature engineering methods ranging from network embedding to heuristic graph structure.
    
    \item Finally, we conduct comprehensive experiments on the competition datasets, which show the effectiveness of our proposal, and exhaustive ablation studies also show the improvement of each kind of feature.
\end{itemize}
Our source code are publicly available on GitHub\footnote{https://github.com/im0qianqian/WSDM2022TGP-AntGraph}.
% Code is publicly available at https://github.com/im0qianqian/WSDM2022TGP-AntGraph.

\section{Datasets}
\label{section:datasets}
In this section, we focus on the exploratory  of datasets provided by the competition, and an in-depth analysis is presented, followed by the detailed introduction of evaluation metrics.

\begin{table*}[t]
  \centering
  \caption{The analysis of the existence of same edges in the initial test set.}
  \label{tab:edge_analysis}
  \begin{tabular}{@{}clllllll@{}}
  \toprule
                             & \multicolumn{1}{c}{Description}   & \multicolumn{1}{c}{\begin{tabular}[c]{@{}c@{}}Total\end{tabular}} & \multicolumn{1}{c}{Exist in graph} & \multicolumn{1}{c}{\begin{tabular}[c]{@{}c@{}}Exist in graph\\ label = 1\end{tabular}} & \multicolumn{1}{c}{\begin{tabular}[c]{@{}c@{}}Exist in graph\\ label = 0\end{tabular}} & \multicolumn{1}{c}{\begin{tabular}[c]{@{}c@{}}Not exist\\ label = 1\end{tabular}} & \multicolumn{1}{c}{\begin{tabular}[c]{@{}c@{}}Not exist\\ label = 0\end{tabular}} \\ \midrule
  \multirow{2}{*}{Dataset A} & $w.o.$ edge type             & \multirow{2}{*}{8197}                                                       & 7354 (89.72\%)                     & 3333 (40.66\%)                                                                         & 4021 (49.05\%)                                                                         & 183 (2.23\%)                                                                      & 660 (8.05\%)                                                                      \\
                             & $w.i.$ edge type &                                                                             & 5886 (71.81\%)                     & 2755 (33.61\%)                                                                         & 3131 (38.20\%)                                                                         & 761 (9.28\%)                                                                      & 1550 (18.91\%)                                                                    \\ \midrule
  \multirow{2}{*}{Dataset B} & $w.o.$ edge type             & \multirow{2}{*}{3863}                                                       & 3195 (82.71\%)                     & 2123 (54.96\%)                                                                         & 1072 (27.75\%)                                                                         & 128 (3.31\%)                                                                      & 540 (13.98\%)                                                                     \\
                             & $w.i.$ edge type &                                                                             & 2612 (67.62\%)                     & 1685 (43.62\%)                                                                         & 927 (24.00\%)                                                                          & 566 (14.65\%)                                                                     & 685 (17.73\%)                                                                     \\ \bottomrule
  \end{tabular}
  \end{table*}
  
\begin{table}[t]
  \centering
  \caption{The performance \wrt AUC of our native strategy compared to the baseline model provided by the sponsor on both initial and intermediate (Inter.) test set.}
  \label{tab:simple_strategy}
  \setlength{\tabcolsep}{1.0mm}{
  \begin{tabular}{@{}clcc@{}}
  \toprule
                             &  Method   & Initial test & Inter. test\\ 
\midrule
  \multirow{3}{*}{Dataset A} & Baseline model             & 0.5110   &  0.5026  \\
                             & Naive strategy ($w.o.$ edge type)             & 0.5428   & 0.5432    \\
                             & Naive strategy ($w.i.$ edge type) & 0.5597   & 0.5687    \\ 
  \midrule
  \multirow{3}{*}{Dataset B} & Baseline model             & 0.5100   &  0.5026  \\
                             &  Naive strategy ($w.o.$ edge type)             & 0.6391   & 0.8655    \\
                             & Native strategy ($w.i.$ edge type) & 0.5867   & 0.8059    \\ 
  \bottomrule
  \end{tabular}}
  \end{table}
  
\subsection{A Brief Description}
The competition expects participants to adopt a single model (hyperparameters can vary) that works well on two kinds of data simultaneously, and thus correspondingly provides two representative large-scale temporal graph datasets. 
\begin{itemize}
    \item \textbf{Dataset A} characterizes a dynamic event graph with entities as nodes and different types of events as edges. Each node maybe associated with rich features if available, and except for the edge types, no any other information is available for edges.
    
    \item \textbf{Dataset B} characterizes a user-item graph with users and items as nodes and different types of interactions as edges. Each edge is associated with rich features if available, and no feature information is available for nodes. Noting that the sponsor treat the user-item graph as a bipartite graph. For convenience, we try to convert this graph to an undirected multi-relation graph through shifting item ids. In particular, we perform the above operation by adding the sum of $1$ and the maximum value of user ids (denotes as $Offset_u$) for each original item id, as follows:
    \begin{equation}
   node\_id=\left\{
\begin{aligned}
node\_id & & {if\ an\ user}\\
 node\_id + Offset_u   & & {if\ an\ item}
\end{aligned} \right. 
\end{equation}

% \begin{equation}
% \label{eq:dst_id_offset}
%   node\_id = dst\_id + \max(src\_id) + 1
% \end{equation}
    
\end{itemize}
Since the competition asks participants to predict whether an edge will exist between two nodes within a given time span, instead of a single timestamp in the graph, \textbf{a start and an end timestamp} is respectively given for each query in test stage. Also, for each dataset, the sponsor provides a train set, an initial test set, an intermediate test set and a final test set, and the labels of intermediate test set and final test set are still not available at present. It is worthwhile to note that \textbf{only the performance of the model in the final test set determines the ranking of the competition}.
In summary, we detailed all necessary statistics of two datasets in Table~ \ref{tab:description_of_datasets}.

\subsection{Data Analysis}
Generally, an inspiring data analysis could shed some light on the model design, which plays a vital role in various data mining tasks. Based on the  originally provided data (\ie the train set and the initial test set), we perform a series of detailed data analysis as follows:
\begin{itemize}[leftmargin=*]
    \item \textbf{The existence of same edges in the test}. 
    We firstly analyze whether edges of the initial test set have already existed in the original graph. In this analysis, timestamps are not taken into consideration. As shown in Table~\ref{tab:edge_analysis}, we observe that the original graph contains most of edges in the initial test for both datasets, especially when the edge type is ignored. Surprisingly, we also find that approximately half of edges (\ie 
 40.66\% $v.s.$ 49.05\% without edge type and 33.61\% $v.s.$ 38.20\% with edge type)  existed in the graph keep the same labels for Dataset A, while about three quarters of edges (\ie 
 54.96\% $v.s.$ 27.75\% without edge type and 43.62\% $v.s.$ 24.00\% with edge type)  existed in the graph keep the same labels for Dataset B. 
 
    Following aforementioned observations, we are curious about the performance of the most naive strategy that just predict the existence of each edge via its existence in the original graphs. We present corresponding results in Table \ref{tab:simple_strategy}, and find that the naive strategy achieve a more competitive performance than baseline model provided by the sponsor. \emph{It indicates the crucial importance of first-order relationship for the task}. 
    
    \item \textbf{Optimal performance without consideration of timestamps}. 
    Secondly, we also explore theoretical upper-bounds on performance without temporal information. We select the data with the same node pair in the initial set, and then calculate the mode or mean value for all labels as the prediction result of these data. Experiments show that the model can still achieve a good performance, as shown in Table \ref{tab:explore_the_maxium_auc}.
\end{itemize}

% After doing EDA (Exploratory Data Analysis), we have obtained a lot of useful and interesting information.

% Firstly, we analyze whether the edge of the dev data exists in the graph without considering the query time.
% As shown in Table \ref{tab:edge_analysis}, the edges of most samples exist in the graph.da
% The edges of a small number of positive samples do not exist in the graph, which indicates that the task may also need to consider link prediction.
% {\itshape Group by} represents the basis used in query.

% 其次，通过实验我们发现只需校验样本中的边是否存在于图上便可获得相比于 baseline 更高的成绩。
% Secondly, through experiments, we find that we can get higher results than baseline\footnote{https://github.com/dglai/WSDM2022-Challenge} by checking whether the edges in the sample exist on the graph (shown in Table \ref{tab:simple_strategy}).

% Finally, based on the previous experiment, we want to know the maximum AUC that the model can achieve in the dataset without time features.
% We aggregate according to the (src\_id, dst\_id) or (src\_id, edge\_type, dst\_id), and then take the mode or mean value of the label column as the prediction result of the sample.
% Experiments show that the model can still achieve a good performance even without time features, as shown in Table \ref{tab:explore_the_maxium_auc}.
\begin{table}[t]
  \centering
  \caption{Explore the maximum AUC without temporal information.}
  \label{tab:explore_the_maxium_auc}
  \begin{tabular}{@{}clcc@{}}
  \toprule
                             & \multicolumn{1}{c}{Description}   & \begin{tabular}[c]{@{}c@{}}Initial test\\ (mode)\end{tabular} & \begin{tabular}[c]{@{}c@{}}Initial test\\ (mean)\end{tabular} \\ \midrule
  \multirow{2}{*}{Dataset A} & node pair ($w.o.$ edge type)             & 0.9040                                                   & 0.9776                                                   \\
                             & node pair ($w.i.$ edge type) & 0.9900                                                   & 0.9997                                                   \\ \midrule
  \multirow{2}{*}{Dataset B} & node pair ($w.o.$ edge type)             & 0.8946                                                   & 0.9795                                                   \\
                             & node pair ($w.i.$ edge type) & 0.9147                                                   & 0.9875                                                   \\ \bottomrule
  \end{tabular}
  \end{table}

\subsection{Evaluation Metrics}
This competition uses Area Under ROC (AUC) as the evaluation metric for both tasks. Intuitively, the two task have different difficulties, and sacrificing one task to do well on the another is not expected. Therefore, the competition further adopt the average of T-scores as the ranking basis for encouraging the model to perform well on different tasks. The formal definition is introduced as follows:
% Since the two tasks have different difficulties, and in order to encourage the model to perform well on different tasks, rather than sacrificing one task to do well on the other. Therefore, the competition uses the average of T-scores as the ranking basis, as shown in Eq. \ref{eq:tscore}-\ref{eq:average_of_tscore}.
\begin{equation}
  \label{eq:tscore}
  Tscore = \frac{AUC - {\rm mean}(AUC)}{{\rm std}(AUC)} * 0.1 + 0.5
\end{equation}
\begin{equation}
  \label{eq:average_of_tscore}
  {\rm AverageOfTscore} = \frac{TScore_A + TScore_B}{2}
\end{equation}
where ${\rm mean}(AUC)$ and ${\rm std}(AUC)$ represents the mean and standard deviation of AUC of all participants. Clearly, an larger average of T-scores means a better performance.

\section{Methodology}
\label{section:methodology}
In this section, we introduce our complete solution for large-scale temporal graph link prediction task, which consists of \emph{train data construction} component, \emph{feature engineering} component and downstream \emph{model training} component. In the following, we will zoom into each well designed component.

%Figure~\ref{fig:1} present the overall framework and pipeline of our solution, which consists of \emph{train data construction} component, \emph{feature engineering} component and downstream \emph{model training} component. In the following, we will zoom into each well designed component.
% In this section, we introduce our framework for Large scale temporal graph link prediction.
% First, we introduce our strategy of sampling raw data and constructing training data. Then, we introduce the work of Feature Engineering for this task. Finally, we introduce several downstream models.
% An overall framework and processing pipeline of our solution is showed in Figure \ref{fig:1}.
% \begin{figure}[ht]
%   \centering
%   \includegraphics[scale=0.40]{./res/pdf/pipeline.pdf}
%   \caption{The processing pipeline of {\modelName}.}
%   \label{fig:1}
% \end{figure}

\subsection{Train Data Construction}
\label{section:train_data_construction}
As mentioned above, the goal of this competition is to predict whether an edge will exist between two nodes within a given \textbf{time span}, whereas each edge in the provided graphs is only associated with \textbf{a single timestamp}. Hence, the inconsistent problem between training and testing severely threatens the generalization of models. In addition, previous data analysis has concluded that this task may not benefit from involving timestamps, therefore, we construct the train data without timestamps as follows:
\subsubsection{Negative sampling}
For efficient training, we adopt the shuffling based sampling strategy to sample negative instance in batch, rather then the whole node set. Moreover, the timestamps are ignored in our negative sampling process. In particular, our negative sampling process is detailed is follows:
i) We denotes  edges in the original graphs as the positive instance set, consisting of source nodes, target nodes and relations. 
ii) We only keep source nodes unchanged, and randomly shuffle target nodes and relations to generate the negative instance set.
iii) We combine the above positive and negative instance set, and uniformly sample a certain number of instances to construct the final train set.

% \subsubsection{Negative sampling}
% This competition provides a heterogeneous graph for each task separately for training,
% but its test task is to query the results at a given time period.
% Obviously, the inconsistency between training and testing tasks will cause trouble to the model.
% Therefore, we construct new training data by sampling based on the original graph.
% The method is as follows:
% \begin{itemize}
%   \item Keep the structure information (src\_id, edge\_type, dst\_id) of the original graph unchanged, add a new label column and set it to ``1'' to represent the positive samples.
%   \item Keep the src\_id column unchanged, randomly shuffle the dst\_id, edge\_type columns, and set the label to ``0'' to represent negative samples.
%   \item Combine positive and negative samples, and then uniformly sample a specified number of samples as new training data.
% \end{itemize}

\subsubsection{Removing redundant features}
Firstly, we remove all time related features, including timestamp, start time, end time. Moreover, we remove the edge features for Dataset B, since these featues are not avaiable for most of edges, \ie the non-empty ratio is 6.67\%.
% For dataset A and B, after our experiments, it is difficult for us to ensure that the distribution of train set and test set is similar when introducing time features.
% A large difference in the distribution of train set and test set is negative for the learning of the model.
% In addition, Section \ref{section:datasets} we also explore the maximum AUC that the model can achieve without time features.
% Therefore, in order to ensure that the model can work better, we remove all time features, including timestamp, start\_time, end\_time.
% Similarly, for dataset B, since edge\_feat are difficult to generate, and the non-empty ratio of edge\_feat is 6.67\%, we also delete this feature.

% \subsubsection{Offset dst\_id}
% As described in Section \ref{section:datasets}, graph B is a bipartite graph consisting of user nodes and item nodes.
% When converting this graph to an undirected homogeneous graph, in order to distinguish user nodes from item nodes.
% We offset all item node ids by a value so that all dst\_id is greater than src\_id (as shown in Eq. \ref{eq:dst_id_offset}).
% \begin{equation}
%   \label{eq:dst_id_offset}
%   dst\_id = dst\_id + \max(src\_id) + 1
% \end{equation}

\subsection{Feature Engineering}
\label{section:feature_engineering}
\subsubsection{LINE embedding}
As concluded in the previous data analysis, the first-order relation is of crucially importance in our link prediction task. In order to capture such deep correlation between nodes in an more fine-grained manner, we introduce the LINE embedding, an effective and  efficient graph learning framework arbitrary graphs (undirected, directed, and/or weighted). In particular, LINE is carefully designed  preserves both the first-order and second-order proximities, which is suitable to our scenarios to capture co-occurrence relation. On the other hands, several heterogeneous ~\cite{hu2019adversarial} and knowledge ~\cite{sun2018rotate,trouillon2016complex,bordes2013translating} graph  representation based methods are also promising ways for learning powerful representations, whereas the LNIE experimentally achieves the best performance, shown in following experiment part.
% In the constructed train set, each sample is an edge whose information is (src\_id, edge\_type, dst\_id).
% In addition to that, there is a label indicating whether the edge actually exists.
% Obviously, this cannot express the information transfer between the event nodes,
% nor can it express whether there is a further association between the user node and the item node.
% For example, if a user purchases two items at the same time,
% there may be a collocation relationship between the two items.
% If a user buys one of the two items,
% there may be a competing relationship between the two items.
% In order to capture the deeper association between nodes, we introduce LINE\cite{tang2015line} embedding.
% LINE is an algorithm for representing large-scale network nodes as low-dimensional vectors.
% It works for any kind of graph, whether directed or undirected, weighted or unweighted.
% LINE can also learn both local and global structural features on the graph.
% This provides extremely useful graph structure information for model training.

\subsubsection{Node crossing features}
After we obtain the representations for each node in graphs, we construct crossing features to further reveal the correlation of each node pair. Specifically, we calculate the similarity of node pairs \wrt LINE embeddings in  datasets as the node crossing features. Given a node pair $(u, v)$ with corresponding embedding $e_u$ and $e_v$, the similarity is calculated through the cosine operation (\ie ${e}_u \cdot {e}_v / {||{e}_u|| \times ||{e}_v||}$) and the dot product (\ie ${e}_u \cdot {e}_v$):
% \begin{itemize}
%   \item Cosine operation: ${e}_u \cdot {e}_v / {||{e}_u|| \times ||{e}_v||}$
%   \item Dot product: ${e}_u \cdot {e}_v$
% \end{itemize}
% After we obtain the representations for each node in graphs, we construct crossing features to further reveal the correlation of each node pair. Specifically, we calculate the similarity of node pairs \wrt LINE embeddings in  datasets as the node crossing features. Given a node pair $(u, v)$ with corresponding embedding $e_u$ and $e_v$, the similarity is calculated with following two ways:
% \begin{itemize}
%   \item Cosine operation: ${e}_u \cdot {e}_v / {||{e}_u|| \times ||{e}_v||}$
%   \item Dot product: ${e}_u \cdot {e}_v$
% \end{itemize}

% In the previous step, we used LINE\cite{tang2015line} to generate representations for each node in the graph.
% At this point, the two nodes in each sample have obtained LINE embedding respectively.
% We call it $LE_{src}$ and $LE_{dst}$.
% As analyzed above,
% the dev data contains some positive samples that did not exist in the original graph.
% This suggests that the association between nodes may not only depend on whether there is an entity edge between two nodes,
% but also on the similarity between two nodes.
% Therefore, we compute the similarity of the LINE embeddings of the two nodes in each sample and use it as a new feature.
% There are two methods of node similarity calculation used in our solution:
% \begin{itemize}
%   \item cosine similarity: $\frac{LE_{src} \cdot LE_{dst}}{||LE_{src}|| \times ||LE_{dst}||}$
%   \item dot product: $LE_{src} \cdot LE_{dst}$
% \end{itemize}

\subsubsection{Subgraph features}

In addition, we also added the following statistical features based on the graph structure to well help downstream model capture high-order information:

\begin{itemize}[leftmargin=*]
     \item Unitary feature \wrt individual nodes:
      i) The degree of this node;
      ii) The number of different nodes adjacent to this node;
      iii) The number of different edge types adjacent to this node.
  \item Binary information \wrt node pairs:
    i) The number of one hop paths between two nodes;
    ii) The number of two hop paths between two nodes;
    iii) The number of different edge types between two nodes.
  \item Ternary information  \wrt node pairs and edge types:
      The number of occurrences of this triplet.
\end{itemize}

% \begin{enumerate}
%   \item Unitary feature \wrt individual nodes
%     \begin{itemize}
%       \item The degree of this node.
%       \item The number of different nodes adjacent to this node.
%       \item The number of different edge types adjacent to this node.
%     \end{itemize}
%   \item Binary information \wrt node pairs
%   \begin{itemize}
%     \item The number of one hop paths between two nodes.
%     \item The number of two hop paths between two nodes.
%     \item The number of different edge types between two nodes.
%   \end{itemize}
%   \item Ternary information  \wrt node pairs and edge types
%     \begin{itemize}
%       \item The number of occurrences of this triplet.
%     \end{itemize}
% \end{enumerate}

\begin{table*}[ht]
  \centering
  \caption{Overall experimental results of different methods in two datasets.}
  \label{tab:experimental_result}
  \begin{tabular}{@{}lllll@{}}
  \toprule
  \multicolumn{1}{c}{\multirow{2}{*}{Model}} & \multicolumn{2}{c}{Dataset A} & \multicolumn{2}{c}{Dataset B} \\
  \multicolumn{1}{c}{}                       & Initial AUC       & Inter. AUC      & Initial AUC       & Inter. AUC      \\ \midrule
  baseline                                  & 0.5110        & 0.5026   & 0.5100        & 0.5026             \\
  % ge, A: initial dotproduct 0.5351596838476849 cossim 0.5348426116372458, mid dotproduct 0.5706748708827452 cossim 0.574120548120422
  % ge, B: initial dotproduct 0.5245700008708564 cossim 0.5252264502239423, mid dotproduct 0.49849271587841815 cossim 0.4943246935408994
  DeepWalk~\cite{perozzi2014deepwalk}        & 0.5352        & 0.5707        &  0.5246             & 0.4985             \\
  % ge, B: initial dotproduct 0.5009477453086745 cossim 0.49285319014543305, mid dotproduct 0.5066712027726541 cossim 0.5456745735546362
  % Node2Vec~\cite{grover2016node2vec}         &               &               &  \red{0.4929}             & \red{0.5457}             \\
  % ampligraph, k=200, epoch=20, A, 0.5173750832098097, 0.5016490050013421
  % ampligraph, k=200, epoch=20, B, 0.4924034313947041, 0.486427903335301
  % KnowledgeGraphEmbedding, 0.51821, 0.56143, 0.63888, 0.89025
  TransE~\cite{bordes2013translating}         & 0.5182       &   0.5614      & 0.6389        & 0.8903       \\
  % KnowledgeGraphEmbedding, 0.53154, 0.57358, 0.63225, 0.89811
  RotatE~\cite{sun2018rotate}                & 0.5315        & 0.5736        & 0.6323        & 0.8981              \\
  % KnowledgeGraphEmbedding, 0.55135, 0.58206, 0.63593, 0.90139
  ComplEx~\cite{trouillon2016complex}        & 0.5514        &  0.5821       & 0.6359        & 0.9014       \\
  LINE~\cite{tang2015line}                   & 0.6072        & 0.6320        & 0.6425        & 0.8905        \\
  % ge, A: initial dotproduct 0.46002453702049706 cossim 0.45777127977720306, mid dot product 0.5006262961390419 cossim 0.49871835405870807
  % ge, B: initial 
  % Struc2Vec~\cite{ribeiro2017struc2vec}       & 0.4600       &  0.5006       &         &         \\
  \midrule
  catboost (raw input data)                  & \underline{0.6045}        & \underline{0.6222}        &  \underline{0.5545}             &  \underline{0.5869}             \\
  + LINE embedding                           & 0.6377 (+5.49\%)        & 0.6540 (+5.11\%)       & 0.6399 (+15.40\%)       &  0.9013 (+53.57\%)            \\
  + Subgraph features                        & 0.6611 (+9.36\%)       & \textbf{0.6673 (+7.25\%)}       &  0.5861 (+5.70\%)        &   0.7561 (+28.83\%)            \\
  + LINE embedding + Subgraph features       & 0.6619 (+9.50\%)       & 0.6659 (+7.02\%)       &  0.6368 (+14.84\%)       &  0.8978 (+52.97\%)             \\
  + LINE embedding + Node crossing features  & 0.6573 (+8.73\%)       & \textbf{0.6673 (+7.25\%)}       &  \textbf{0.6504 (+17.29\%)}       &  0.9001 (+53.37\%)           \\
  + All (submitted version)                  & \textbf{0.6657 (+10.12\%)}       & 0.6671 (+7.22\%)       & 0.6459 (+16.48\%)       & \textbf{0.9028 (+53.83\%)}       \\
  \bottomrule
  \end{tabular}
  \end{table*}
  
\subsection{Catboost Model}
The link prediction task can be easily formulated as a binary classification problem based on the extracted features from each (source node, relation, target) triple. On the other hands, gradient boosting has prove its powerful capability in various applications for classification. Recently,  catboost~\cite{prokhorenkova2018catboost} has gained increasing popularity and attention due to its  fast processing speed and high prediction performance. We feed the train data (see Section \ref{section:train_data_construction}) as well as abundant features (see Section \ref{section:feature_engineering}) into catboost model, and then utilize the produced scores as the final predictions.
% Gradient boosting is a powerful algorithm especially for binary-classifying problem.
% Recently, catboost\cite{prokhorenkova2018catboost} has gained increased popularity and attention due to their advantages of fast processing speed and high prediction performance. 
% In our solution, we use catboost model as the downstream classification model.
% Catboost receives the result of feature engineering as shown in Figure \ref{fig:1} as input and outputs the score of the model for each sample.
% Finally, the scores of all samples are the submitted data.
% The performance of the model is the result of AUC calculated by the score and label column.

% \subsubsection{NN Model}
% \red{to be continue...}

\section{Experiments}

\subsection{Overall Performance}
\noindent \textbf{Performance in the leaderbord}.
We present the results of top five teams from the learderboard in Table~\ref{tab:top_5_results}. We observation that our solution achieve the best performance in Dataset A and competitive performance in Dataset B. As the best performance achived \wrt the final ranking metric further indicating that our solution works well on both kinds of data simultaneously.

\noindent \textbf{Compare to baselines}.
We compare our methods with other 6 methods, including the official baseline\footnote{https://github.com/dglai/WSDM2022-Challenge} and several classic network embedding methods, \ie  DeepWalk~\cite{perozzi2014deepwalk},
TransE~\cite{bordes2013translating},
RotatE~\cite{sun2018rotate},
% Node2Vec~\cite{grover2016node2vec},
LINE~\cite{tang2015line},
% Struc2Vec~\cite{ribeiro2017struc2vec}
and ComplEx~\cite{trouillon2016complex}.
The experimental results in Table~\ref{tab:experimental_result} shows that our solution outperforms all baselines by a considerable margin.

Overall, both of observations verfy the effectiveness of our proposal.

% According to the final ranking metric, we win the
% first place in the temporal link prediction competition, demonstrating the effectiveness of our solution.

% which show the effectiveness of our solution. In particular, we achieve the best performance in Dataset A, and 

% Table \ref{tab:top_5_results} shows the results of the top five teams from the leaderboard.
% Our system achieves state-of-the-art AUC in Dataset A.
% Meanwhile, experimental results also show that our system performs well in solving Dataset B. 
\begin{table}[t]
  \centering
  \caption{Top five results in the final learderboard.}
  \label{tab:top_5_results}
  \begin{tabular}{@{}ccccc@{}}
  \toprule
  Rank       & Team name                & \begin{tabular}[c]{@{}c@{}}Dataset A\\ Final AUC\end{tabular} & \begin{tabular}[c]{@{}c@{}}Dataset B\\ Final AUC\end{tabular} & \begin{tabular}[c]{@{}c@{}}Average of\\ T/100\end{tabular} \\ \midrule
  \textbf{1} & \textbf{AntGraph(Ours)} & \textbf{0.666001}                                             & 0.901961                                             & \textbf{0.630737}                                          \\
  2          & nothing here             & 0.662482                                                      & \textbf{0.906923}                                                      & 0.628942                                                   \\
  3          & NodeInGraph              & 0.627821                                                      & 0.865567                                                      & 0.585137                                                   \\
  4          & We can {[}mask{]}!       & 0.603621                                                      & 0.898232                                                      & 0.572372                                                   \\
  5          & IDEAS Lab UT             & 0.605264                                                      & 0.873949                                                      & 0.566849                                                   \\ \bottomrule
  \end{tabular}
  \end{table}

% \subsection{Compare to baselines}
% We compare our methods with other 6 methods, including the official baseline\footnote{https://github.com/dglai/WSDM2022-Challenge} and several classic network embedding methods, \ie  DeepWalk~\cite{perozzi2014deepwalk},
% RotatE~\cite{sun2018rotate},
% % Node2Vec~\cite{grover2016node2vec},
% LINE~\cite{tang2015line},
% ComplEx~\cite{trouillon2016complex},
% % Struc2Vec~\cite{ribeiro2017struc2vec}
% and TransE~\cite{bordes2013translating}.

\subsection{Ablation studies}
In this section, we perform a series of ablation studies to analyses the impact of  kinds of features proposed in Section \ref{section:feature_engineering}, including LINE embedding,  node crossing features and subgraph features. We summarize the comparison results in Table \ref{tab:experimental_result}, and we have following observations:
i) All extracted feature help base model achieve better performance, and the best performance is yielded in most cases with the incorporation of all features.
ii)  Involving LINE features show a greater improvement than models involving subgraph features in Dataset B, while opposite trend is observed in Dataset A. An intuitive explanation is that Dataset B is much sparser than Dataset A, and thus subgraph structure are hardly exploited in Dataset B.
Overall, compared to the base model only using raw feature, our final submitted model achieve a relative improvement of 7.22\% on dataset A and 53.83\% on dataset B, respectively.

\section{Conclusion}
This paper describes our solution for WSDM 2022 Challenge - Temporal Link Prediction.
For this task, we design a novel negative sampling strategy, and combined with data analysis to delete redundant information.
We introduce the LINE embedding to provide local and global features of graph.
At the same time, we design node crossing features and subgraph features.
In the end, our team {\itshape AntGraph} was ranked the 1st place on the final leaderboard.
% Even though our model achieves the best performance,
% it is still our disadvantage that temporal features are not considered.
% In the future, we will pay more attention to the modeling of time series information on the graph.
% Meanwhile, we will further explore to model the fusion of node information on the graph through the graph neural network, and enhance the graph-based reasoning ability of the model.

% \begin{acks}
% acks
% \end{acks}

\bibliographystyle{ACM-Reference-Format}
\bibliography{main}

\appendix

\end{document}